# Neuroplasticity and Psychedelics:

## A comprehensive examination of classic and non-classic compounds in pre and clinical models

Claudio Agnorelli[1,2], Meg Spriggs[1], Kate Godfrey[1], Gabriela Sawicka[1], Bettina Bohl[3], Hannah Douglass[1], Andrea Fagiolini[2], Hashemi Parastoo[3], Robin Carhart-Harris[1,4], David Nutt[1], and David Erritzoe[1]

1 Centre for Psychedelic Research, Division of Psychiatry, Department of Brain Science, Imperial College of London, UK.

2 Unit of Psychiatry, Department of Molecular and Developmental Medicine, University of Siena, Italy

3 Department of Bioengineering, Imperial College of London, UK

4 Departments of Neurology and Psychiatry, Carhart-Harris Lab, University of California San Francisco, San Francisco, CA, USA

**Abstract**

Neuroplasticity, the ability of the nervous system to adapt throughout an organism's lifespan, offers potential as both a biomarker and treatment target for neuropsychiatric conditions. Psychedelics, a burgeoning category of drugs, are increasingly prominent in psychiatric research, prompting inquiries into their mechanisms of action. Distinguishing themselves from traditional medications, psychedelics demonstrate rapid and enduring therapeutic effects after a single or few administrations, believed to stem from their neuroplasticity-enhancing properties. This review examines how classic psychedelics (e.g., LSD, psilocybin, N,N-DMT) and non-classic psychedelics (e.g., ketamine, MDMA) influence neuroplasticity. Drawing from preclinical and clinical studies, we explore the molecular, structural, and functional changes triggered by these agents. Animal studies suggest psychedelics induce heightened sensitivity of the nervous system to environmental stimuli (meta-plasticity), re-opening developmental windows for long-term structural changes (hyper-plasticity), with implications for mood and behavior. Translating these findings to humans faces challenges due to limitations in current imaging techniques. Nonetheless, promising new directions for human research are emerging, including the employment of novel positron-emission tomography (PET) radioligands, non-invasive brain stimulation methods, and multimodal approaches.  By elucidating the interplay between psychedelics and neuroplasticity, this review informs the development of targeted interventions for neuropsychiatric disorders and advances understanding of psychedelics' therapeutic potential.

*"And one finds oneself, to put it all in perspective, in a situation where fifty different onomatopoeias, simultaneous, contradictory, and each constantly changing, would be the most faithful expression of it."* Henri Michaux, Misérable Miracle, 1956

1. **Introduction**

The ability of the nervous tissue to modify over the course of an organism's life is referred to as "neuroplasticity". Modern definitions of neuroplasticity generally distinguish between structural and functional plasticity (Figure 1). Though closely interconnected and not mutually exclusive, structural plasticity involves the physical shaping of the nervous system, whereas functional plasticity pertains to alterations in synaptic communication. (Cohen et al., 2017). Such definitions, which are detailed at the level of individual neurons or networks of neurons, are primarily based on preclinical research in animal models. Less is known about how these processes can be captured in clinical research. However, new neuroimaging methods are being developed to close this translational gap, providing a chance to evaluate the preclinical findings in humans.

In evolutionary terms, neuroplasticity can lead to gain or loss of functions which can be viewed as either adaptive or maladaptive. Additionally, many neuronal structures and functions have a limited time window to optimally develop. These windows are known as **critical periods** of plasticity, and the majority tend to close before adulthood (Ismail et al., 2017). Therefore, the context and timing of neuroplasticity alterations are critical factors in determining their significance. Despite this, neuroplasticity is gaining traction both as a potential biomarker of neuropsychiatric illnesses and as a target for therapeutics.

Psychedelics, from the Greek words psychḗ 'soul, mind' and dēleín 'to manifest', are a large family of drugs that are currently being (re-)explored as potential treatments for mental health disorders (See Box 1 for discussion on nomenclature). The most studied compounds are the classic psychedelics, such as psilocybin, lysergic acid diethylamide (LSD), N,N-Dimethyltryptamine (N,N-DMT), and 5-methoxy-N,N-dimethyltryptamine (5-Meo-DMT), and the non-classic psychedelics, such as ketamine and 3,4-Methylenedioxymethamphetamine (MDMA). Sufficient doses of psychedelics reliably induce a transient altered state of consciousness, followed by long-lasting changes in mood, personality, and behavior (Knudsen, 2023). Such changes and their underlying biology have been found to endure far longer than the presence of the drug in the organism (Garcia-Romeu et al., 2016). Evidence from contemporary clinical trials has shown that the symptom relief produced by psychedelic substances surfaces after one or a few treatment sessions yet can last for an extended period without re-exposure to the drug (Nutt et al., 2020). The novelty of psychedelics compared to available mental health medications lies in their rapid onset of action and long-lasting therapeutic properties.

The non-classic psychedelic ketamine was one of the first compounds whose potential as a treatment for mental illness was investigated in modern trials. At the start of the millennium, it was found that a single psychedelic dose of ketamine administration resulted in significant antidepressant effects within a few hours (Berman et al., 2000). According to the most recent research, ketamine's therapeutic effects as a standalone treatment for depression and bipolar disorder peak after 24 hours and can last up to a week afterward, with a few additional doses prolonging the effects up to 4 weeks after the last treatment (Walsh et al., 2022). MDMA has been mainly investigated for its potential as a complement to psychotherapy in treating post-traumatic stress disorder (PTSD), with long-term reductions in symptoms being reported. Results from existing clinical trials indicate that with 2 to 3 active sessions, the therapeutic effects were still present 1 to 2 months after the final MDMA-assisted session (Mitchell et al., 2021; Mithoefer et al., 2019). Classic psychedelics have been researched for various mental health conditions, including anxiety, depression, obsessive-compulsive disorder, and substance abuse disorders (Andersen et al., 2021). Rapid and sizable effect sizes enduring for follow-up durations ranging from months to years after the conclusion of therapy are remarkable aspects of these investigations. It's interesting to note that the long-lasting effects of classic psychedelics go beyond symptom relief and include psychological changes such as increased nature connection (Kettner et al., 2019), sexual functioning (Barba et al., 2024), mindfulness (Madsen et al., 2020; Smigielski et al., 2019), openness (Erritzoe et al., 2018; Lebedev et al., 2016; MacLean et al., 2011; Madsen et al., 2020; Weiss et al., 2021a), extraversion (Erritzoe et al., 2018; Weiss et al., 2021a), agreeableness (Weiss et al., 2021a, 2021b), spirituality (Griffiths et al., 2011, 2008, 2006), well-being (Carhart-Harris et al., 2012; Griffiths et al., 2008, 2006; Perkins et al., 2022), and social functioning (Kettner et al., 2021; Weiss et al., 2021b).

There is growing evidence of defective neuroplasticity in illnesses where psychedelics have shown to be effective, including mood disorders and addiction. These include structural and cellular alterations, such as neuronal loss and synaptic dysfunction, in cortico-limbic brain regions controlling mood and emotions (Kalivas and O'Brien, 2008; Manji et al., 2000). Thus, a framework is emerging around the idea that psychedelics may rectify these alterations by inducing a heightened state of plasticity which provides a window of opportunity for therapeutical intervention with enduring efficacy (Aleksandrova and Phillips, 2021; Castrén and Antila, 2017; Kavalali and Monteggia, 2020).

It must be noted that psychedelics are not the only plasticity-enhancing pharmacological agents, as currently prescribed antidepressants, like selective serotonin-reuptake inhibitors (SSRIs), have also been demonstrated to modify neuroplasticity when exerting their therapeutic effects (Castrén and Antila, 2017; Tardito et al., 2006). Also, heightened state of plasticity are not necessarily a positive thing. For example, plasticity within reward circuity has been identified as a mechanism of addiction to compounds such as cocaine and the plasticity-enhancing capacity of psychedelics may underlie aspects of negative effects such as the hallucinogen persisting perception disorder (HPPD) (Rossi et al., 2023; Thomas et al., 2008). Additionally, current understandings of neuroplasticity lack the depth and precision required to fully harness the potential of plasticity-enhancing agents in therapy. In the case of psychedelics, there has been little consideration for how the parameters of the plasticity window induced by the different compounds may vary, e.g., regarding timeframe and intensity. Nevertheless, it appears that a mechanistically distinct ability to promote neuroplasticity that is unique in both intensity and kind may account for the difference in effect sizes and speed of response to psychedelics compared with currently available mental health medications. Although many reviews have explored the plasticity-promoting effects of psychedelics (Aleksandrova and Phillips, 2021; Calder and Hasler, 2023; Inserra et al., 2021; Liao et al., 2024; Olson, 2022; Slocum et al., 2022) , they predominantly focus on preclinical evidence, often neglecting translational validity and comprehensive cross-compound analysis. The purpose of this manuscript is to compare the available evidence about the impact of classic psychedelics, ketamine, and MDMA on neuroplasticity in both animal and human research. This review will emphasize the translational value of preclinical research to human studies, with a focus on identifying specific target mechanisms to establish a framework for guiding future investigations.

The studies included in this review have been selected to cover the effects of ketamine, classic psychedelics, and MDMA, on neuroplasticity, therefore only discussing findings relative to measures that are thought to quantify specific and well-characterized neuroplastic processes. A summary of all studies referenced in this review is provided in Supplementary Table 1. In describing the results, the distinction between in vivo and in vitro findings refers to when the drug was administered, either in the living organism or sampled tissue, respectively. Given the high variability of dosages and routes of administration used by the various studies, doses will be referred to as: "low dose" when they approximate ranges that do not induce alterations in consciousness and haven't yet shown convincing evidence of rapid therapeutic efficacy in humans (e.g., micro-doses or mini doses of psychedelic drugs); "medium dose", when they approximate ranges that induce alterations in consciousness and show evidence of therapeutic efficacy in humans (e.g., active psychedelic doses of classic psychedelic drugs and sub-anesthetic, psychedelic doses of ketamine). In the case of ketamine we use "high dose" where they approximate the anesthetic doses of ketamine. For classic psychedelics and MDMA, medium (e.g. 10-15 mg psilocybin) and high dosages (e.g., 25 mg of psilocybin) are both referred to as "medium dose" for easier comparison with the ketamine literature.

> *Box 1.* Terminology of psychedelics:
>
> The question of what term accurately describes the large class of mind-altering compounds examined in this review is up for debate, and there is still a lack of consensus. We use the term "psychedelic" in this article but make a distinction between the classic psychedelics, strictly defined as being agonists at the 5-HT2AR, and the non-classic psychedelics, which include the NMDAR antagonist ketamine and the entactogen MDMA. The rationale for grouping different compounds under the name of psychedelics lies in the common property of such compounds to induce an altered state of consciousness often with insights into the person's self-understanding and with potential as an adjuvant for psychotherapy.

> Recently, the word "psychoplastogens" has been coined to describe substances that cause rapid and long-lasting neuroplastic alterations after a single or a few doses, such as ketamine, classic psychedelics, and MDMA. However, the term is only used in preclinical research and is also applied to substances that do not produce psychedelic-like effects in animal models.

## 2. Molecular mechanisms of action

Ketamine is an arylcyclohexylamine, here classified as a non-classic psychedelic, with a marked dose-dependent profile of psychoactive effects, generally referred to as dissociative. In fact, at medium doses ketamine-induced dissociative effects are psychedelic-like, transitioning into anaesthesia as dosage increases. Ketamine acts primarily as an antagonist of the N-methyl-D-aspartate receptor (NMDAR) expressed on neurons releasing glutamate and γ-Aminobutyric acid (GABA). In clinical settings, ketamine is frequently given intravenously in a racemic mixture of its two enantiomers, namely S (+)- and R (-)-ketamine, though both given separately are also psychedelic and the S (+) enantiomer -esketamine – is now licensed as a medicine. The subjective effects of ketamine are mainly linked to its glutamatergic mechanism of action in cortical regions. Like other NMDAR antagonists, ketamine binds to the intra-channel phencyclidine site of the NMDAR, sitting on the GluN2 subunit. Ketamine binding results in a decrease of the channel opening and a decrease in the amplification of the response to repeated stimulation (Mion and Villevieille, 2013). In frontocortical regions, the administration of ketamine produces an increase in glutamate release. These effects seem paradoxical considering the NMDAR antagonistic properties of the drug, and different mechanistic theories have been proposed. One of the main hypotheses proposes that at sub-anesthetic doses, ketamine preferentially blocks NMDARs expressed on inhibitory GABAergic interneurons, thus disinhibiting the glutamatergic neurons they regulate. The rise in glutamate brought on by ketamine activates postsynaptic α-amino-3-hydroxy-5-methyl-4-isoxazolepropionic acid receptor (AMPAR), releasing Brain-derived neurotrophic factor (BDNF) which activates the Mammalian target of rapamycin (mTOR) via agonism to the Tropomyosin receptor kinase B (TrkB) (Figure 1). In this manner, an autoregulatory feedback loop is initiated, whereby activation of BDNF activates its own transcription (Zanos et al., 2018).

Classic psychedelics are defined by their direct activity on the serotonin (5-HT) system, producing alteration of the perception of reality and the sense of self. Members of this family of substances belong to three main chemotypes, namely phenethylamines (e.g., mescaline and synthetic substituted phenethylamines like DOI and 25C-NBOMe), ergolines (e.g., LSD), and tryptamines (e.g., psilocybin, N, N-DMT, and 5-Meo-DMT). Despite the pharmacological diversity within classic psychedelics, agonism of the 5-HT receptor type 2A (5-HT2AR) appears to be the primary mechanism via which these drugs' subjective effects are mediated (Nichols, 2016). The 5-HT2AR is a canonical G protein-coupled receptor and is highly conserved across species, apart from a single amino acid which differs between humans and other mammals with substantial effect on the affinity and efficacy of a variety of 5-HT2AR agonists (Slocum et al., 2022). The 5-HT2AR is densely expressed in neocortical regions involved in cognition, perception, sensorimotor gating, and mood (Ettrup et al., 2014). In particular, glutamatergic pyramidal neurons in layer V of the prefrontal cortex exhibit a high density of 5-HT2ARs in their apical dendrites, both within the membrane and inside the cell (Vargas et al., 2023). When stimulated by classic psychedelics, 5-HT2AR-expressing glutamatergic neurons initiate a glutamate cycling involving the same AMPAR-BDNF-TrkB-mTOR past-synaptic pathway described for ketamine (Aleksandrova and Phillips, 2021). The release of glutamate via 5-HT2AR agonists is mediated by a changed excitability of glutamatergic neurons and the direct association of 5HT2AR with glutamate receptors. Further 5HT2AR agonists activate three main intracellular signalling cascades, namely the Gq/11, the β-arrestin-2, and the arachidonic acid (AA) pathways (Figure 1; other pathways also exist, for a detailed review see (Slocum et al., 2022)). These pathways activate general cell activity, survival, morphology and maturation pathways by directly impacting gene expression. In particular, the Gq/11 protein complex is triggered by 5-HT2AR stimulation, leading to Gα detachment from Gβγ subunits and exchange of guanosine diphosphate (GDP) for guanosine triphosphate (GTP). GTP-bound Gα activates phospholipase C (PLC), causing breakdown of phosphatidylinositol bisphosphate (PIP2) into inositol trisphosphate (IP3) and diacylglycerol (DAG). IP3 releases intracellular Ca2+, activating CaMKII, while DAG activates PKC, which phosphorylates ERK and CREB, influencing gene expression. Gβ/γ subunits activate phospholipase A2 (PLA), releasing arachidonic acid (AA) and modulating gene expression. Classic psychedelics also engage β-arrestin-2, initially known for 5-HT2AR internalization but later found to induce neuronal activation and gene expression via Src, Akt, and ERK phosphorylation. Virtually all psychedelics trigger those pathways to induce neuroplasticity, as also do some non-psychedelic analogues and the endogenous ligand 5-HT, but there is significant compound-specific variability in the relative signalling potencies and subjective effects those molecules produce. The mechanistic model proposed to account for the therapeutic action of psychedelic show partial overlaps with ketamine, where the long-lasting behavioral and psychological changes are supported by an increased state of neuroplasticity.

MDMA is an amphetamine derivative but is here regarded as non-classic psychedelic because even though it shows some low affinity for 5-HT2AR (Nash et al., 1994) its subjective and neuroimaging effects are very different from classic 5-HT2AR acting psychedelics (Carhart-Harris et al., 2015; Roseman et al., 2014). MDMA shares structural similarities with the classic psychedelic mescaline, but due to its strong prosocial effects, MDMA is frequently referred to as an "entactogen". MDMA can interact, either directly or indirectly, with the 5-HT transporter (SERT), the dopamine (DA) transporter (DAT), the norepinephrine (NE) transporter (NET) and directly stimulates 5-HT release and, to a lesser extent, also DA and NE release (5-HT pathway shown in Figure 1) (Costa and Gołembiowska, 2022). Evidence suggests that the psychological effects of MDMA in humans are largely driven by its impact on 5-HT, with the drug's empathogenic properties linked to downstream modulation of oxytocin levels (Liechti et al., 2000; Nardou et al., 2019).

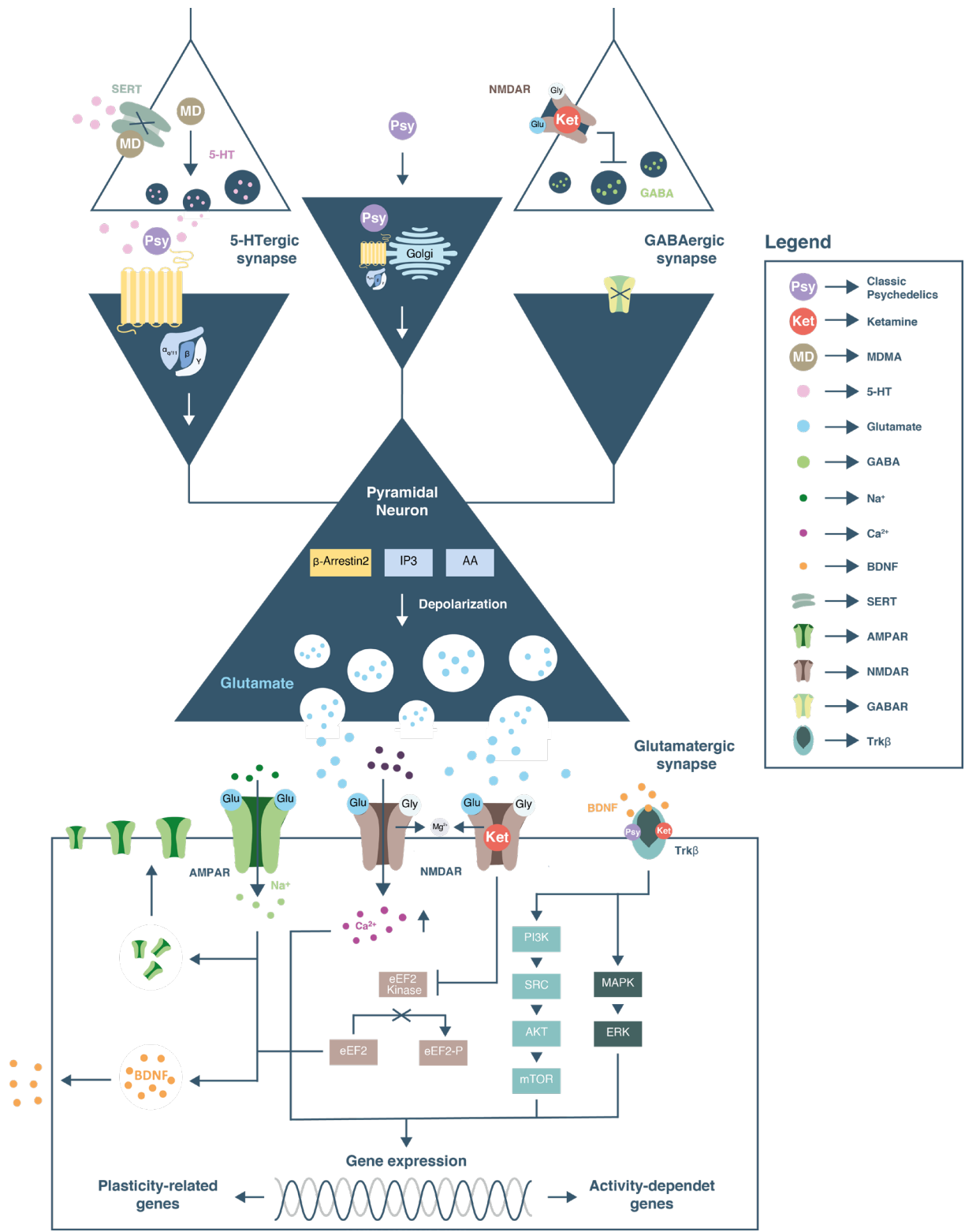

*Figure 1: The acute cellular mechanism of classic psychedelics, ketamine and MDMA.*

### 3. Structural plasticity

Structural plasticity encompasses all those processes that regulate the morphological re-modelling of the brain. This phenomenon includes the formation of new neurons (i.e., neurogenesis) or their death (i.e., neuronal apoptosis), the increases in neuronal structural complexity (i.e., neuritogenesis) or its decrease (i.e., neuronal atrophy), the increases in the number of connections with other neurons (i.e., synaptogenesis) or its decrease (i.e., synaptic pruning) (Bernardinelli et al., 2014; Götz and Huttner, 2005). Neuritogenesis is a process whereby the neurons undergo a remodeling of their appendixes, resulting in the formation of new dendrites (dendritogenesis), new spines (spinogenesis), or axonal elongation. Synaptogenesis refers to the formation of new connections between neurons and their maturation into functional synapses (Bernardinelli et al., 2014). Pharmacological interventions that directly enhance these processes can be referred to as mediators of hyper-plasticity, while interventions that modulate the thresholds of neurons to undergo structural plasticity can be referred to as mediators of structural meta-plasticity (Figure 2) (Nardou et al., 2023).

In animal models, structural plasticity can be studied with various biological techniques. Common approaches to studying neurogenesis involve the use of specific molecular markers for developmental stages, from stem cells to mature neurons, in isolated neural tissue (La Rosa et al., 2020). Neuritogenesis and synaptogenesis are investigated by morphological assays or immunohistochemical staining of membrane-associated synaptic proteins in vitro or, more recently, via two-photon microscopy, which can be used also in the living organism (Bernardinelli et al., 2014). In humans, few techniques have been developed for the study of structural plasticity in the living brain. Neurogenesis and neuritogenesis can be studied via structural magnetic resonance (MR) and diffusion tensor imaging (DTI) to quantify grey and white brain matter, respectively (May, 2011). Recently, [$^{11}$C]-UCBJ, a novel positron emission tomography (PET) radioligand has been developed as the first in vivo human measure of synaptic plasticity. The ligand binds to the synaptic vesicle glycoprotein 2A (SV2A), an integral membrane protein located in presynaptic vesicle membranes. It is expressed ubiquitously and homogenously across synapses in the normal brain and has been validated for labelling synaptic plasticity in primates and humans (Bajjalieh et al., 1994; Finnema et al., 2016; Nabulsi et al., 2016). The concentration of SV2A is believed to mainly reflect the density of synapses in a given region of the brain, making it suitable for imaging of structural plasticity (Nabulsi et al., 2016). However, increases in SV2A could also result from a rise in the number of vesicles at the presynaptic terminal, potentially due to Hebbian learning, or from changes in the number of SV2A molecules per vesicle (Rossi et al., 2022). Therefore, while [$^{11}$C]-UCBJ is among the most precise tracers currently available for in vivo imaging of synaptic plasticity, it remains uncertain whether it actually reflects structural or functional neuroplastic processes, or both.

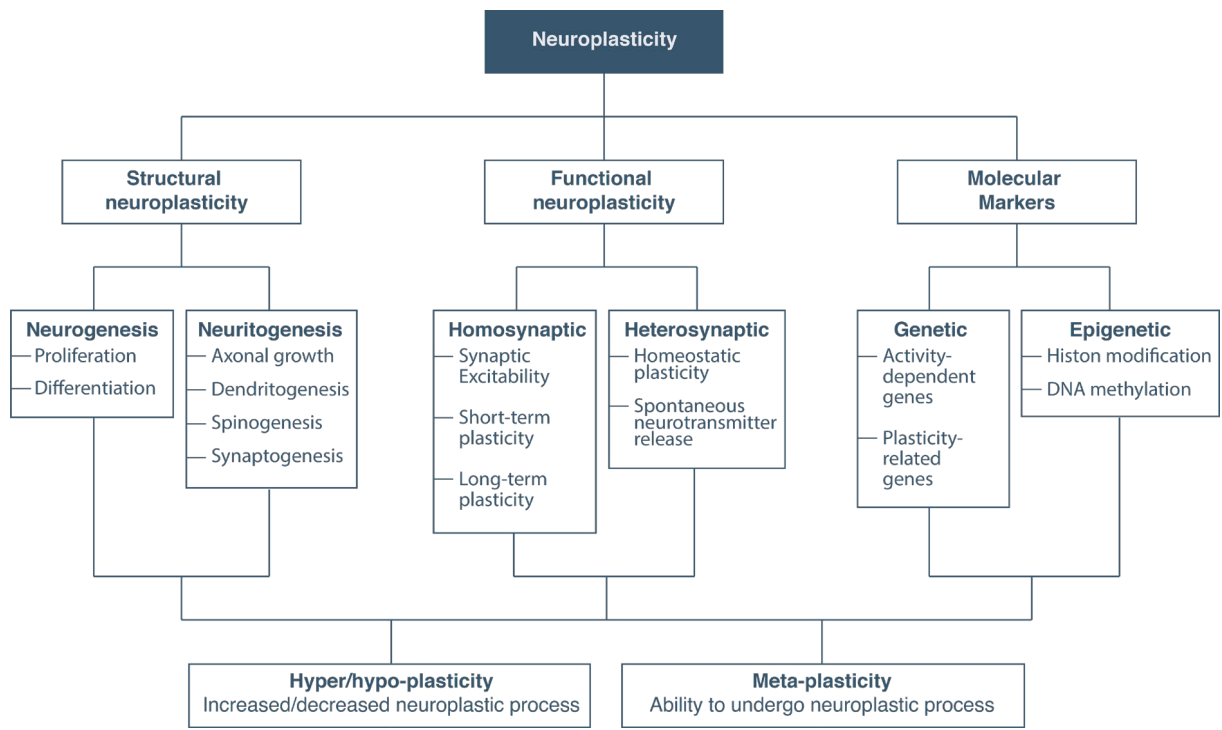

*Figure 1: Taxonomy of neuroplastic processes.*

**3.1. Ketamine**
**3.1.1. Preclinical**

In vitro, treatment of animal-derived brain tissue with ketamine has been found to affect structural neuoplasticity. Various concentrations of ketamine promoted the proliferation phase of cultivated rat neural stem and progenitor cells (NSPCs) 3 weeks after exposure. However, the high concentrations, translating to anesthetic doses in humans, inhibited the proliferative phase of neurogenesis 24 hours after treatment (Dong et al., 2012). Different studies have shown that rat-derived neurons exposed to medium concentrations of ketamine, corresponding to dissociative sub-anesthetic doses in humans, increased spinogenesis, dendritogenesis, dendritic arbor complexity, and soma size after 72 hours, with lowest concentrations and short periods of exposure having the highest efficacy (Cavalleri et al., 2018; Ly et al., 2021, 2018; Zhang et al., 2020). Structural modifications were found to depend on an AMPAR-BDNF-TrkB-mTOR signalling by a few studies (Cavalleri et al., 2018; Ly et al., 2021, 2018) and the phosphorylation of CRMP2, a downstream target of mTOR, by one study (Zhang et al., 2020).

In vivo administration of a single medium dose of ketamine has been shown to promote structural plasticity in a large number of studies. Accelerated progenitor cell differentiation into newborn neurons and maturation of new neurons in the dentate gyrus (DG) of mice was seen 24 and 48 hours after ketamine administration. Interestingly, ketamine promotion of late neurogenesis and activity of DG adult-born immature granule neurons was associated with the long-term antidepressant-like effects of the drug (Ma et al., 2017; Rawat et al., 2022; Yamada and Jinno, 2019). One of the studies showed that these effects were mediated by TrkB-dependent ERK pathway activation (Ma et al., 2017). A few studies have reported that ketamine increased spine density in the PFC at 24 hours following administration, an effect associated with the emergence of an antidepressant-like response, that lasted up to 2 weeks after a single sub-anaesthetic dose (Li et al., 2010; Phoumthipphavong et al., 2016; Ruddy et al., 2015). Li et al. showed that the effect was mediated by mTOR (Li et al., 2010). Another study found an increase in the total density of dendritic spines in rat cerebral cortex 6 hours after injection of ketamine, which depended on the phosphorylation of CRMP2 (Zhang et al., 2020). Recently, dual laser two-photon glutamate uncaging and imaging were used to resolve the temporal dynamics of the ketamine-induced spinogenesis in the mPFC layer 5 pyramidal neurons. Administration of a single medium dose of ketamine to mice enhanced glutamate-evoked spinogenesis (i.e., neuron's likelihood to form new spines) 2 and 4 hours after treatment, temporally matching the emergence of ketamine's therapeutic effects. However, 12 hours after ketamine was administered, such "plasticity potential" decreased back to baseline levels. In parallel, the actual increase in dendritic spine density manifested only after 12 hours following ketamine and lasted for 72 hours post-treatment. This timing sequence where ketamine first enhances plasticity potential implies that alterations in cellular properties, influencing the propensity to form new spines, play a role in the gradual rise of spine density following ketamine treatment. In an uncontrollable stress paradigm the probability of glutamate evoking spinogenesis decreased relative to the baseline, with ketamine treatment restoring the baseline potential for plasticity (Wu et al., 2021). Interestingly, these effects were found in Drd1-expressing mPFC neurons to be dependent upon the activation by DA neurons of the ventral tegmental area (VTA) (Wu et al., 2021). In rodent models of depression, a single medium dose of ketamine reversed the depressive-like behaviour and PFC spine density deficit at 3, 12, and 24 hours post-injection. These effects were not due to modification of elimination rates but to an increase in spine formation rates (Moda-Sava et al., 2019; Sarkar and Kabbaj, 2016). A similar result was obtained with (S)-ketamine, which induced spinogenesis in the hippocampus 1 hour post-injection, and increased excitatory synaptogenesis 1 to 7 days after treatment (Treccani et al., 2019). However, the effects observed in models of depression were not observed in the wild-type controls. One study found no increase in spine density in the hippocampus 3 hours post-injection of a medium dose of ketamine (Widman et al., 2018). Importantly, prolonged exposure to high anesthetic doses of ketamine, administration during development, and/or chronic administration of the drug have been found to induce neuronal degeneration and reduce neurogenesis and cortical spine density (Brown et al., 2015; Huang et al., 2016; Ruddy et al., 2015; Zou et al., 2009). However, one study showed that pre-treatment with a single high dose of ketamine before prolonged exposure to an anaesthetic dose was able to prevent neuronal cell death, an effect mediated by upregulation of Activity-Dependent Neuroprotective Protein (ADNP) (Brown et al., 2015).

### 3.1.2. Human

In human studies, direct evidence for the effect of ketamine on structural plasticity is still scarce. In vitro, human-derived stem cells treated with high concentrations of ketamine show increased neural proliferation but also neural apoptosis after 24 hours, while various concentrations increase dendritic arborization and soma size measured after 72 hours (Bai et al., 2013; Cavalleri et al., 2018). In vivo, a recent study employed the novel PET radioligand 11CUCB-J in depressed patients, patients with trauma, and healthy subjects. At 24 hours after a single medium dose of ketamine, no significant changes in the concentration of SV2A were detected in all groups. The same null finding was observed in macaques imaged at 24 hours, 1 week, and 4-6 weeks after ketamine administration. A post-hoc exploratory analysis showed that ketamine produced a significant increase in SV2A density only in those depressed patients showing the lowest SV2A density at baseline (Holmes et al., 2022). Several factors could have contributed to such results. First, the difference in neuroplastic changes induced by ketamine between depression endophenotypes could point to a genuine inter-individual variability in drug response rooted in neurophysiological differences. This view would be supported by some of the previously reported preclinical evidence (Bærentzen et al., 2024; Treccani et al., 2019) and by findings in population with schizophrenia using [$^{11}$C]-UCBJ, whereby significant differences compared to healthy controls are only manifest in patients with severe symptomology (Onwordi et al., 2023). However, an effect of regression to the mean cannot be ruled out given the post-hoc and exploratory nature of the result. Another possibility is that repeated or higher dosages of ketamine might be required to produce a detectable effect at the group level, as suggested by recent evidence obtained with SSRIs. Johansen et al. observed that chronic SSRI intake over 3–5 weeks led to a significant increase in SV2A density, but only in patients who underwent a longer duration of escitalopram treatment (Johansen et al., 2023). It is worth reminding that while we placed the [$^{11}$C]-UCBJ finding in the structural plasticity section, consistent with the main interpretation of it indexing synaptic density, contribution of other processes, such as Hebbian plasticity and vesicle trafficking cannot be disambiguated. DTI can also be used to quantify increases in synaptic density via decreases in mean diffusivity within the grey matter, albeit indirectly. Using this technique, a study (Kopelman et al., 2023) found that MDD patients treated with a medium dose of ketamine show no significant main effect on mean diffusivity 24 hours post-infusion. Relative reductions in mean diffusivity in the left Brodmann area 10 and left amygdala, representing putative increased plasticity in these regions, predicted greater improvement in depression scores in patients receiving ketamine. However, in the hippocampus, the results were in the opposite direction, with higher mean diffusivity predicting greater improvement in depression scores (Kopelman et al., 2023). With structural MRI it was shown that ketamine reduced the volume of the left nucleus accumbens (NAc) but increased left hippocampal volume in depressed patients 24 hours post-dose. However, the results were only significant for patients achieving remission (Abdallah et al., 2017). Repeated injections of medium doses of ketamine resulted in an increase in MRI volumes of the right hippocampus and left amygdala in MDD patients 24 hours after the last administration. However, at baseline, the volume in those regions was smaller compared to healthy individuals (Zhou et al., 2020). Another study found that repeated administration of oral ketamine tablets resulted in a bilateral volume increase in the putamen, thalamus, caudate, NAc, and the periaqueductal grey in patients with chronic suicidality (Gallay et al., 2021). Using tensor-based morphometry, it was found that a single dose of ketamine increased volume in several brain regions in MDD patients, but not in healthy controls. Within those regions, the inferior frontal gyrus was found to be smaller in MDD compared with controls at baseline, and 24 hours after ketamine administration this abnormality was normalized, in association with antidepressant effects. However, another study found the volume of the left lateral orbitofrontal cortex was significantly reduced in MDD after ketamine administration (Dai et al., 2020). A few studies have shown that when compared to drug-naive controls, long-term ketamine abusers showed brain changes, such as bilateral frontal and left temporoparietal white matter reductions and decreased grey matter volume over the right insula, left dorsolateral PFC, left orbitofrontal cortex, and left inferior parietal cortex (Hung et al., 2020; Liao et al., 2011, 2010; Wang et al., 2013).

### 3.2. Classic psychedelics

### 3.2.1. Preclinical

Classic psychedelics have been shown to upregulate structural plasticity in animal models both in vitro and in vivo. First in vitro results are shown, followed by in vivo results, both organised according to the specific mechanism (such as neurogenesis, dendritogenesis, etc.) and chronological proximity to the intervention. In vitro cultivated neurospheres, neural progenitor-enriched cellular 3D-structures, isolated from the DG of adult mice chronically exposed to low concentrations of N,N-DMT, corresponding to microdoses in humans, showed an increased proliferation 24 hours after treatment and differentiation 72 hours after treatment (Morales-Garcia et al., 2020). Chronic exposure to medium concentrations of DOI, LSD and, N,N-DMT, corresponding to psychoactive doses in humans, to rat-derived primary cortical neurons produced an increase in both the complexity of dendritic arbors and the total length of the branches 72 hours after drug exposure. At comparable concentrations, psychedelics were found to have higher potency in promoting dendritogenesis as compared to ketamine. Treatment with DOI, LSD, and N,N-DMT also increased the number of immature spines and synaptic density on the dendrites (Dunlap et al., 2020; Ly et al., 2018). Even brief exposure of 15 min to medium concentrations of LSD was sufficient to initiate dendritogenesis, with an effect size comparable to the effects observed with 1 hour of treatment followed by a 72 hours growth period. Though overall for spinogenesis and synaptogenesis, a 6 hours treatment period followed by 18 hours of growth appeared to be the most efficient. Like ketamine, the observed psychedelic-induced neuroplastic effects were dependent on sustained AMPAR and mTOR signalling. Importantly, longer stimulation periods produced smaller changes in both dendrite and spine growth (Ly et al., 2021). In a recent discovery, Vargas et al. (Vargas et al., 2023) showed that the ability of psychedelics to induce dendritogenesis relies on the activation of 5-HT2AR pools located **intracellularly** rather than on membrane surface. They showed that while molecules that permeate the neuronal plasma membranes such as N,N-DMT and psilocin promote dendritogenesis, their impermeable chemical analogues as well as 5-HT, can only promote plasticity if experimentally allowed to enter the cell. Also, activation of transmembrane 5-HT2AR alone was found to be not sufficient to induce structural plasticity, but activation of intracellular 5-HTAR was required. Thus, the physio-chemical properties of the different 5-HT2AR agonists would determine their ability to induce dendritogenesis, with highly polar molecules, which have more difficulties in permeating lipid membranes such as the blood-brain barrier and cellular membranes, showing lower neuroplastic properties. The finding was recently confirmed by a computational model of various 5-HT2AR agonists (Palmisano et al., 2024). Vargas et al. also showed that a mice model with ectopic post-synaptic expression of SERT in the medial PFC of mice, allowing 5-HT to access and activate intracellular 5-HT2ARs, have increased spine density and antidepressant-like effects 24 hours after pharmacological stimulation of 5-HT release (Vargas et al., 2023). In contrast with Vargas et al., however, a recent study showed that psychedelics can promote structural neuroplasticity via a 5-HT2AR-independent mechanism. Moliner et al. (Moliner et al., 2023) observed that rodent-derived cortical neurons treated with equal low concentrations of LSD and psilocin, the active metabolite of psilocybin, showed increases in dendritic arbor complexity and spinogenesis at 24 hours mediated by TrkB expressed by the neurons. The effect persisted after the blockage of 5-HT2ARs. The study suggests a novel mechanism of action of antidepressants, such as classic psychedelics, ketamine, and SSRIs, whereby the allosteric binding to TrkB would promote the dimerization of the receptor and localization near raft-like synaptic cellular membranes, increasing TrkB sensitivity to BDNF binding (i.e., structural meta-plasticity). In this study, psychedelics showed a higher affinity for allosteric binding to TrkB than did SSRIs (1,000-fold greater), potentially accounting for the higher efficacy of the former in inducing neuroplasticity via such a mechanism (Moliner et al., 2023). In fact, SSRIs-mediated increases in structural plasticity requires chronic treatment regimens to manifest (Hajszan et al., 2005; Johansen et al., 2023; Ogelman et al., 2024; Seo et al., 2014). While some residues involved in TrkB binding are shared between psychedelics and SSRIs, the binding modes are different, with psychedelics forming more stable interactions with TrkB. However, future research is needed to reconcile the findings of Moliner et al. with those of Vargas et al. regarding the role of the 5-HT2A receptor using comparable experimental methodologies.

In vivo, a single intracerebral injection of a low concentration of 5-MeO-DMT increased neural proliferation 12 hours after treatment and increased neural differentiation 3 days after treatment, in the DG of adult mice (Lima da Cruz et al., 2018). In contrast, a single injection of a medium dose of psilocybin decreased the number of newly formed neurons, 2 weeks after treatment, an effect not observed with lower doses. Also, lower doses facilitated the extinction of conditioned fear response, while the higher dose did not affect this behavioral output (Catlow et al., 2013). Mice treated with repetitive administration of medium doses of N,N-DMT showed enhanced proliferation and migration of neural precursors in the subgranular zone of the DG immediately after treatment. Also, there was an enhanced migration of newly formed neurons in the granular cell layer of the hippocampus after 21 days from the last dose. In blockage experiments, the effects of N,N-DMT on neurogenesis were found to be mediated by the sigma-1 receptor (S1R), rather than 5-HT2AR, and were associated with an increase in performance in behavioral tests of episodic memory (Morales-Garcia et al., 2020). Notably, the action of N,N-DMT on S1R was previously reported to mediate protective effects against hypoxia in human-derived cortical neurons and microglia-like immune cells (Szabo et al., 2016). In addition to their in vitro findings, Moliner et al. showed that a single medium dose of LSD enhances long-term neuronal survival of granule cells in the DG of mice 4 weeks after administration, in a TrkB-dependent manner (Moliner et al., 2023). Other studies found that a single medium dose of N,N-DMT, or DOI produce an increase in the density of dendritic spines of cortical neurons in the PFC of adult rats 24 hours after dosing. These effects were dependent on 5-HT2AR and mTOR activation and for DOI were associated with the acceleration of fear extinction learning (de la Fuente Revenga et al., 2021; Ly et al., 2018). Similar results were obtained in pigs after a single injection of a medium dose of psilocybin, showing higher synaptic density as measured via SV2A (measured via [$^3$H]-UCBJ post-mortem autoradiography) in the hippocampus 24 hours post-dose and higher SV2A in the hippocampus and PFC 7 days post-dose (Raval et al., 2021). Adult mice exposed to chronic restraint stress showed a reduction in the number of dendritic spines in the medial PFC, and repeated administration of medium doses of LSD during the last 7 days of the stress paradigm reversed the structural deficit at 24 hours after the last dose, an effect associated with a reversal of anxiety- and depressive-like behaviors. In control mice, LSD administration produced an increase in spinogenesis but did not produce behavioral effects (De Gregorio et al., 2022). Interestingly, one study found that intermittent administration of microdoses of N,N-DMT did not produce significant changes in PFC spine density in male rats but reduced the number of spines in female animals 24 hours after the treatment. However, the same dosing regimen produced an antidepressant-like phenotype and enhanced fear extinction learning (Cameron et al., 2019). Shao et al. used tracked apical dendritic spines growth in the medial PFC of mice and found that a single medium dose of psilocybin induced a significant elevation in spine density, an increase in the width of spine heads, and higher spine protrusion lengths 7 days after treatment. Remarkably, a fraction of the psilocybin-evoked new spines remained persistent for a month after injection. The observed neuroplastic effects were due to an initial boost of enhanced spine formation and were not blocked by partial blockage of the 5-HT2AR (Shao et al., 2021). The effect of a single injection of 5-Meo-DMT was assessed with the same paradigm, resulting in a rapid increase in spine density which persisted for a month following injection. Also in this case, the increase in spine density observed with the drug was due to an initial boost of spine formation rate (specifically during the first and third day after treatment), with no effect observed in spine elimination rates (Jefferson et al., 2023). Consistently, a single intracerebral injection of 5-MeO-DMT increased the number of branches in dendrite trees and the dendritic complexity of newborn granule cells in the DG of adult mice 21 days after treatment (Lima da Cruz et al., 2018).

### 3.2.2. Human

To date, no direct evidence has been collected on the effect of classic psychedelics on structural plasticity in humans. The only published data comes from an MRI study conducted on long-term users of ayahuasca (a preparation whose active principle is N,N-DMT) as compared to controls. Thinning was observed in the ayahuasca-using group in six cortical areas: the middle frontal gyrus, the inferior frontal gyrus, the precuneus, the superior frontal gyrus, the posterior cingulate cortex, and the superior occipital gyrus. On the contrary, thickening was found in the precentral gyrus and the anterior cingulate cortex (ACC) (Bouso et al., 2015). In a report currently under review, Lyons and colleagues showed that a single medium dose of psilocybin decreased axial diffusivity bilaterally in prefrontal-subcortical tracts (i.e., increased structural plasticity) measured via DTI at 1 month following drug administration in psychedelic-naïve healthy subjects (Lyons et al., 2024).

### 3.3. MDMA
#### 3.3.1. Preclinical

Only a few studies have measured the effect of MDMA effects on structural plasticity, with high variability of methods and experimental design. In vitro exposure to MDMA of rat-derived tumour cells treated showed a dose-dependent increase in cell death 24 hours after treatment (Ball et al., 2009). A later study found that cultured cortical neurons treated with low concentrations of MDMA produce a similar increase in neuritogenesis (i.e., increase in arbor complexity, number of dendritic branches, and the total length of the branches) as ketamine and psychedelics 72 hours after treatment (Ly et al., 2018).

In vivo, different paradigms employing repetitive administration of high doses of MDMA reported increases in apoptosis at 24 hours and 14 days and decreases in neurogenesis at 28 days in rats (Hernández-Rabaza et al., 2006; Renoir et al., 2008; Soleimani Asl et al., 2015). Daily administration of low and high doses of MDMA from day 6 of pregnancy to day 21 after delivery resulted in a reduction of cell proliferation in the DG of the offspring 11 weeks after birth. The high dose alone reduced the survival of newly generated postmitotic cells in the DG of adult female offspring (Cho et al., 2008). Conversely, in another study repeated exposure to high doses of MDMA produced a significant increase in cell proliferation 24 hours after treatment in the DG of adolescent rats. However, there was also a reduction in immature cell survival and neuronal differentiation 2 weeks after exposure (Catlow et al., 2010). Repeated injection of high doses of MDMA produced an increase in spine density in the NAc 4 days after the last dose. Also, there was an increase in apical and basal dendrites in the ACC and prelimbic cortex (Ball et al., 2009). Adolescent rats treated with repeated injections of high doses of MDMA showed a reduction of spine density in the hippocampus after 1 week. However, training in the Morris water maze prior to MDMA treatment resulted in an increase in spine density compared with controls and these neuroplastic effects were associated with the enhancement of both learning and memory processes (Abad et al., 2014). Preliminary results of structural MRI in non-human primates after repetitive administration of MDMA showed an increase in the volume of the frontal cortex, occipital cortex, caudate nucleus, hippocampus, midbrain, and amygdala 60 and 66 months after treatment (Yeh et al., 2022).

#### 3.3.2. Humans

In humans, the only published studies have showed that long-term MDMA users present a reduction in grey matter volume relative to non-users in several brain regions, such as the frontal cortex, temporal cortex, occipital lobe, cerebellum, and pons (Cowan et al., 2006, 2003). However, in these studies participants were often poly-drug users, complicating the assessment of MDMA effects.

### 4. Functional plasticity

Functional plasticity is often subdivided into activity-dependent (or homosynaptic plasticity) and activity-independent (or heterosynaptic plasticity) (Citri and Malenka, 2008). Activity-dependent plasticity is a type of plasticity that depends on positive feedback loops and includes phenomena such as short-term plasticity and long-term plasticity (i.e., 'Hebbian plasticity'). Short-term plasticity includes processes happening in the timeframe of milliseconds to minutes and refers to the phenomenon in which synaptic efficacy changes over time in a way that reflects the history of presynaptic activity. Examples of this type include short-term synaptic facilitation, depression, and habituation (Zucker and Regehr, 2002). Long-term plasticity refers to changes in synaptic strength and efficacy observed in the timeframe of minutes to hours and is thought to be the biological process underlying classical forms of learning and memory. Based on the strength, frequency, and number of pre-synaptic inputs, a synapse can undergo long-term potentiation (LTP) or depression (LTD) (Citri and Malenka, 2008). Activity-independent types of plasticity act at wide timescales and include different forms of homeostatic plasticity such as synaptic scaling and spontaneous neurotransmitter release. Homeostatic plasticity refers to the ability of neurons to adjust synaptic or intrinsic excitability via a negative feedback loop to keep firing rates relatively constant within a network (Kavalali and Monteggia, 2020). When either activity-dependent or independent forms of plasticity are directly enhanced by a pharmacological intervention the result can be termed functional hyper-plasticity, while an intervention that changes the threshold to undergo the various forms of functional synaptic plasticity is referred to as functional meta-plasticity (Figure 2) (Hulme et al., 2014).

Experimentally, functional properties of neurons can be studied via recording brain activity in response to either a sensory or electrical stimulation. It has been found that based on the intensity, pattern, and frequency of stimulation different types of functional plasticity can be induced (Abraham et al., 2002; Bliss and Collingridge, 2013; Dudek and Bear, 1992; Figurov et al., 1996; Mulkey and Malenka, 1992). For example, in animal research a typical paradigm involves the recording of excitatory post-synaptic potentials (EPSP) evoked in a population of neurons after electrical stimulation of afferent pre-synaptic neurons. The EPSP evoked by a single stimulation can be used to measure changes in the excitability of neurons, resulting from activity-dependent or independent forms of plasticity that happened in the synapse (Dudek and Bear, 1992). Also, stimulation can be used to induce short-term or long-term changes. To induce short-term changes in the synapse, a paradigm called paired-pulse stimulation is often used, measuring the modulation of EPSP by two consecutive electrical pulses (Debanne et al., 1996). To induce LTP, a repetitive electrical stimulus is applied at high-frequency, producing an increase in the amplitude of the EPSP between pre- and post-stimulation (Abraham et al., 2002; Figurov et al., 1996). Conversely, to induce LTD, the stimulus is delivered at a low frequency, producing a decrease in the EPSP (Dudek and Bear, 1992; Mulkey and Malenka, 1992). Those induction paradigms can be used to index meta-plasticity, as they measure the ability of the brain to undergo short or long-term changes. Other approaches measure functional re-organization following moto-sensory deprivation, which include paradigms such as the ocular dominance test, where the ocular dominance shift of neurons in the visual cortex is measured after monocular deprivation (Wiesel and Hubel, 1965).

In humans, sensory stimulation inducing evoked-response potential (ERP) in magnetoencephalography (MEG) or electroencephalography (EEG) activity can be used to measure short- and long-term functional neuroplastic changes in humans (see Box 2 for description).

---

Box 2. Measuring functional neuroplasticity in humans:

In humans, the ERP-evoked by somatosensory stimulation or transcranial magnetic stimulation (TMS) while recording MEG or EEG activity can be used to measure cortical excitability (Cornwell et al., 2012; Pascual-Leone et al., 1998). Short-term plasticity can be measured via the mismatch negativity (MMN) responses, the earliest cognitive component of the ERP (Garrido et al., 2009b). The MMN is typically induced using "Oddball" or "Roving MMN" (rMMN) paradigms. Oddball paradigms typically involve embedding a rare, or "deviant" auditory stimulus into a train of frequent or "standard" auditory stimuli. The deviant tone is a divergence from learnt expectations about the sensory input, which leads to an error response in the form of the MMN ERP: a large, fronto-central negativity peaking approximately 100-250 ms post-stimulus. The "explaining away" of prediction error over peri-stimulus time corresponds to perceptual inference and is dependent on sensitivity to context (Garrido et al., 2009a). The roving MMN paradigm (rMMN) builds on this by utilizing sequences of tones, where the first tone in each sequence acts as the deviant. As the stimulus becomes more predictable (over repetitions), prediction error decreases- a pattern known as repetition suppression. This model updating, or perceptual learning, is dependent upon the optimization of the balance between predictions and prediction errors, also known as precision updating or gain control. The rMMN is dependent on short-term synaptic plasticity (though see (Todd et al., 2012) for an MMN paradigm for assessing modulation of priors on the timescale of tens of minutes). Long-term plasticity can be induced via high-frequency sensory stimulations to index meta-plasticity (Clapp et al., 2005; Teyler et al., 2005). The paradigms developed by Teyler et al. were the first non-invasive protocol for inducing long-term changes in humans using sensory stimulation. In the visual version of the paradigm, participants are presented with low-frequency (approximately 1 Hz) presentations of simple visual stimuli such as sine gratings both preceding and following a high-frequency (approximately 9 Hz) presentation during EEG recording. This high-frequency stimulus, referred to as a "tetanus", is somewhat analogous to the high-frequency electrical stimulation used in rodent studies and induces an enhancement of the ERP to subsequent presentations of the stimulus.

Importantly, this enhancement outlives the timescale of short-term adaptation, with post-tetanus experimental blocks typically being conducted up to 40 minutes post-tetanus. Numerous studies employing variations of this procedure have demonstrated an enhancement of the evoked response following tetanization that is consistent with Hebbian mechanisms (see (Sumner et al., 2020c) for a full review). Another technique that has been proven to induce LTP in humans is repetitive TMS stimulation combined with EEG, a paradigm that is similar to sensory-evoked LTP but uses high-frequency repetitive TMS pulses to induce plasticity (Esser et al., 2006).

Data coming from non-invasive neuroimaging techniques in humans are often coupled with computational modeling to build biologically-informed models. The leading computational modeling technique for MMN and vLTP ERP studies to date has been Dynamic Causal Modeling (DCM). DCM is one specific flavor of biologically informed computational modeling that can be used to infer upon the neuronal parameters both within (e.g., intrinsic connectivity) and between (e.g., extrinsic connectivity) modeled neuronal sources that underlie a particular evoked response (Friston et al., 2003). In ERP research, DCMs typically utilize neural-mass models to construct networks of a limited number of sources within a functionally relevant network. The first study to use DCM to explore the networks underlying visual LTP and rMMN was presented by Spriggs et al., in 2018 (Spriggs et al., 2018). The rMMN was modeled using a fronto-temproral network consistent with previous literature (note that this is the same model used in rMMN studies discussed below). The visual LTP network was identified using a hypothesis-driven model comparison approach and included occipital, temporal and, frontal sources. This study demonstrated that visual LTP was associated with the modulation of forward connections, while the MMN was associated with the modulation of reciprocal forward and backward connections within their respective networks. This study provided the first empirical demonstration that differing Hebbian and Predictive Coding mechanisms can be measured using computational modeling.

### 4.1. Ketamine
### 4.1.1. Preclinical

In vitro, rat visual cortex neurons exposed to high concentrations of ketamine showed increased excitability, as indexed via increased amplitude of the baseline response to single-pulse stimulation-evoked EPSPs stimulation, but reduced LTP. Following the washout of ketamine for 30 min, the reduction in induced LTP was normalized (Salami et al., 2000). In hippocampal slice cultures, stimulation of CA3-Schaffer collateral pathway to CA1-stratum radiatum (CA3-CA1 synapses) during exposure to high and low concentrations of ketamine increased excitability only at specific low stimulation frequencies (Narimatsu et al., 2002) while others found dose-dependent inhibition of CA1-evoked EPSPs (Izumi and Zorumski, 2014) or no effect at low concentrations (Graef et al., 2015). High-frequency stimulation (HFS)-induced LTP during ketamine treatment was found to be either impaired (Graef et al., 2015) or unaffected at low concentrations (Izumi and Zorumski, 2014), and inhibited at high concentrations (Nosyreva et al., 2013). In one study, a low concentration of ketamine did not affect HFS-induced LTP but acutely reduced low-frequency stimulation (LFS)-induced LTD. Just after exposure to low concentration of ketamine, there was enhanced excitability for at least 2 hours (Izumi and Zorumski, 2014), and 1 hour following drug washout (Nosyreva et al., 2013). When HFS was administered 1 hour after ketamine washout, LTP was increased by low concentrations of ketamine. When HFS was administered 2 to 4 hours after ketamine washout, however, LTP was inhibited (Izumi and Zorumski, 2014).

Several studies investigated the functional neuroplastic effect of in vivo administration of a single medium dose of ketamine at different time points. One study found that ketamine increased neuronal excitability measured via single-pulse evoked EPSPs in the CA3-CA1 synapse of rat hippocampal slices compared to control 24 hours after drug exposure (Burgdorf et al., 2013). In a depression-like mice model, ketamine reversed the deficit in cortical excitability as indexed via single-pulse-induced NMDA-dependent EPSPs (Yang et al., 2018). Short-term plasticity indexed via paired-pulse-induced facilitation in CA3-CA1 synapses was found to be unaltered at 3 hours post drug administration in wild-type (Widman et al., 2018) and depression-like mice model, but increased at 24 hours in the depression model (Aleksandrova et al., 2020). In wild-type rodents, response to stimulation-induced LTP in CA3-CA1 synapses after ketamine was found to be increased at 3 hours with theta-burst stimulation but not HFS (Aleksandrova et al., 2020; Widman et al., 2018), and increased at 24 hours with combined theta-burst/HFS (Burgdorf et al., 2013) or HFS, and increased at 72 hours with HFS (Graef et al., 2015). In models of depression, the deficit in response to stimulation-induced LTP was found to be reversed by ketamine at 3 and 24 hours with HFS (Aleksandrova et al., 2020; Yang et al., 2018). The neurophysiological effects observed in the depressed model were associated with antidepressant-like response and improvement of hippocampal-dependent long-term spatial memory and contextual fear conditioning (Aleksandrova et al., 2020; Yang et al., 2018). Using monocular deprivation paradigms it was shown that medium doses of ketamine accelerates functional recovery of adult mice's visual cortex, reopening the critical window of developmental plasticity, either after a single dose (Cannarozzo et al., 2023; Grieco et al., 2020) or repeated exposure (Cannarozzo et al., 2023; Casarotto et al., 2021). The effect of ketamine on ocular dominance has been proposed to be mediated by allosteric binding to the TrkB receptor, with a mechanism similar to what has been reported for classic psychedelics and SSRIs (Casarotto et al., 2021; Moliner et al., 2023). Further, it was shown that ketamine disrupts the interaction between TrkB and PTPσ, a tyrosine phosphatase that acts as a receptor for components of perineuronal nets surrounding inhibitory interneurons, a mechanism which might allow for the re-opening of functional critical windows of plasticity (Cannarozzo et al., 2023). In an adult mice model of amblyopia, ketamine injection following 2 weeks of monocular deprivation promoted functional recovery of deprived eye performance in the visual water maze task. These effects were found to be dependent on the expression of the tyrosine kinase receptor ErbB4, target of the neurotrophic factor Neuregulin-1, expressed on V1 inhibitory interneurons. It was found that via downregulation of this NRG1/ErbB4 pathway on inhibitory neurons, ketamine treatment decreases inhibitory inputs to L2/3 excitatory neurons at 1, 24, 48, 72 hours, and 1 week as measured via inhibitory post-synaptic current (IPSC) amplitudes following electrical stimulation in V1 visual cortex. This effect was paralleled by a reduction of the excitatory inputs to inhibitory neurons, but no changes to the excitatory inputs to V1 excitatory neurons. As a result of this cortical disinhibition, the activity of excitatory neurons, measured via two-photon calcium imaging, significantly increased at 24, 48, and 72 hours following ketamine and returned to baseline after 1 week. Also, ketamine treatment increases the visually evoked potential amplitudes in V1 at 24 and 48 hours after treatment (Grieco et al., 2020). In a recent study, Nardou et al. showed that a single medium dose of ketamine re-opens the critical period for social reward learning in adult mice, as measured via the social reward conditioned place preference assay, measured at 48 hours after administration. This behavioral effect was found to be paralleled by meta-plastic changes in Nac, where ketamine was found to potentiate the LTD induced by exposure to oxytocin. No changes in baseline EPSC amplitude or frequency following ketamine treatment were detected in the Nac or medial PFC (Nardou et al., 2023). High doses of ketamine either enhanced (Graef et al., 2015) or unaltered HFS-induced LTP at 24 hours (Ribeiro et al., 2014) post-injection in CA3-CA1 synapses of wild-type mice. No changes in neuronal excitability were detected after a high anesthetic dose of ketamine with single-pulse-evoked EPSP in the dorsal raphe nucleus at 24 hours (Llamosas et al., 2019). Also, neonatal exposure to high doses of ketamine-induced an impairment of LTP evoked by HFS in the DG after 10 weeks, with no difference in single-pulse-evoked EPSCs (Guo et al., 2018).

In summary, the effects of ketamine on neuroplasticity and excitability exhibit complex dose- and time-dependent dynamics across various neural systems. In vitro studies demonstrate variable effects of ketamine on LTP and LTD, but consistent increases in baseline excitability. In vivo, ketamine enhances excitability and LTP in CA3-CA1 synapses us to 24 hours after drug exposure, particularly in depression models, where it reverses deficits in excitability and plasticity, correlating with antidepressant-like behavior and cognitive improvements. Ketamine increases in meta-plasticity also correlate with the reopening of critical periods of plasticity in the visual cortex and for social reward learning, mediated by mechanisms such as neurotrophic TrkB and ErbB4 signaling.

### 4.1.2. Humans

In humans, Sumner et al. used the visual LTP paradigm during EEG in combination with DCM (Box 2) to study the effect of a single injection of a medium dose of ketamine in patients with treatment-resistant

MDD measured 3-4 hours after drug administration. They found that the P2 component of the ERP was significantly more positive in the late post-tetanus block after ketamine compared with the active placebo, showing higher meta-plasticity induced by ketamine (Sumner et al., 2020a). In a sister study, an rMMN task was administered 3-4 hours post ketamine injection to measure repetition suppression of sensory-evoked EEG components. Ketamine was found to increase the negativity of the MMN response to deviant tones, indicating increased sensitivity to prediction error. Also, the source-level results demonstrate a significant main effect of ketamine on the strength of activation in the inferior temporal cortex in response to deviant tones, which showed higher activation post-ketamine (Sumner et al., 2020b). Another line of research investigated the effects of ketamine in treatment-resistant MDD at 6.5 hours with MEG during a tactile stimulation to measure stimulus-evoked somatosensory cortical excitability. This type of stimulation elicited an evoked response approximately after stimulus with a spectral peak in the gamma band over the contralateral hemisphere. Responders showed significantly increased somatosensory cortex gamma band response after ketamine relative to baseline while non-responders showed no change (Cornwell et al., 2012) (Those results were followed-up by a replication study (Nugent et al., 2019)). Another study used MEG and DCM with the same paradigm in both treatment-resistant MDD patients and healthy controls at 6 to 9 hours after a single medium dose of ketamine or placebo. It was found that the NMDA-mediated backward connectivity from the right frontal cortex to the right primary somatosensory cortex after ketamine was higher for patients compared with ketamine-treated controls. Also, there was an increase in NMDA-mediated connectivity in the forward connection from the right somatosensory cortex to the right frontal cortex for healthy subjects following ketamine administration compared to the placebo, but not baseline, within the same group. These results were interpreted as an effect of ketamine-induced NMDA antagonism leading to short-term sensitization of postsynaptic mechanisms, affecting forward and backward NMDA connectivity separately for both MDD subjects and healthy controls (Gilbert et al., 2018).

### 4.2. Classic psychedelics
#### 4.2.1. Preclinical

In rodents, increased spontaneous EPSPs were registered after in vivo administration of a medium dose of N,N-DMT in hippocampal slices or psilocybin in slices of layer 5 pyramidal neurons (Ly et al., 2018; Shao et al., 2021). Injection of psilocybin was also associated with the amelioration of maladaptive behavior induced by uncontrollable stress (Shao et al., 2021). Administration of a medium dose of LSD in rats produced no changes in baseline in vivo local field potential and coherence in the NAc and the infralimbic cortex 24 hours post-dose. However, in vivo cortical excitability indexed via single-pulse stimulation of infralimbic cortex before and 24 hours after LSD resulted in a bigger pre- to post-change in brain activity compared to control within the infralimbic cortex, NAc, and orbitofrontal cortex. In particular, LSD produced stimulation-induced increases in delta coherence and decreases in theta, alpha, and beta coherence and selective delta and high gamma power increases in local filed potentials (Dwiel et al., 2023). A single administration of 5-MeO-DMT in the mice DG produced shorter afterhyperpolarization potential duration and higher action potential threshold induced via patch clamp (i.e., direct electrical current applied through a small glass pipette attached to the cell membrane) in newborn neurons 21 days post-dose compared to controls. Also, treatment with 5-Meo-DMT caused an increase in spontaneous EPSCs in newborn neurons (Lima da Cruz et al., 2018). Repeated administration of medium doses of LSD 24 hours before recording, produced an increase in the mean spontaneous cell firing frequency of dorsal-raphe nucleus 5-HT neurons measured via in vivo single-unit extracellular recordings compared to control mice. In addition, LSD produced an increased coefficient of variation percentage suggesting that repeated LSD administration triggered a more irregular firing activity as opposed to the regular or rhythmic firing activity. Application of a chronic-stress paradigm produced anxiety-like and depressive-like behavior and decreased the spontaneous firing rate activity of 5-HT neurons compared to control, and chronic LSD treatment reversed these effects (De Gregorio et al., 2022). In the work by Nardou et al. it was shown that at medium doses, classic psychedelics re-open the critical window of social reward learning for longer than ketamine, with effects lasting up to 2 weeks for psilocybin and 3 weeks for LSD. In contrast to ketamine, those effects were dependent on 5-HT2AR-mediated signalling. Also, the meta-plastic increase in LTD induced by oxytocin treatment in the NAc was found to last up to 2 weeks after a medium LSD dose, with no changes in baseline EPSP amplitude or frequency observed in the NAc or layer 5 of mPFC (Nardou et al., 2023).

### 4.2.2. Humans

The only study assessing the post-acute effects of a classic psychedelic on functional neuroplasticity in humans was only recently published. The study by Skosnik et al., measured the effect of a single administration of a medium dose of psilocybin, or placebo, in patients with MDD using an auditory-induced LTP EEG paradigm at 24 hours and 2 weeks post dose. In their design, binaural tone pips are used instead of visual grids to induce ERPs via LFS and LTP via tetanic stimulation. The primary outcome measure was evoked theta power before and after auditory tetanus, which is the temporal-spectral equivalent of the N100-P200 ERP complex used by similar paradigms. At both 24 hours and 2 weeks, there was no statistically significant difference in auditory-evoked theta power before and after LTP for either psilocybin or placebo. In a post-hoc analysis, it was observed that overall auditory-evoked theta power averaged across before and after auditory LTP was significantly increased 2 weeks after psilocybin, but not at 24 hours, and this increase correlated with the reduction of depressive symptoms (Skosnik et al., 2023). However, interpreting such result in the context of neuroplasticity is challenging.

## 4.3. MDMA
### 4.3.1. Preclinical

In vitro application of low concentrations of MDMA resulted in increased neural excitability indexed via single-pulse stimulation-evoked EPSP amplitude in CA3-CA1 synapses of rat hippocampal slices 2 hours after washout (Mlinar et al., 2008).

In vivo, daily administration of medium doses of MDMA resulted in a reduction of metaplastic response to theta-burst stimulation-evoked LTP in CA3-CA1 synapses of rat hippocampal slices 24 hours after last the dose. However, there was no difference in paired-pulse facilitation-evoked EPSP in MDMA treated rats compared to control. This dose regime also produced impairments in spatial learning performance (Catlow et al., 2010). Another study found that treatment with repeated medium doses of MDMA increased meta-plastic response to theta-burst stimulation-evoked LTP in CA3-CA1 synapses in rat hippocampal slices measured 1 week after last the dose. No changes in neural excitability measured via single-pulse-evoked fEPSPs were reported between control and MDMA-treated rats (Morini et al., 2011). Similarly to ketamine and the classic psychedelics, it was found that a single medium dose of MDMA re-opens the critical period for social reward learning in adult mice after 6 hours and up to 2 weeks after exposure. Interestingly, this effect was recapitulated by optogenetic stimulation of oxytocin terminals in the NAc. Also, there was a meta-plastic increase in LTD induced by oxytocin treatment in the NAc 48 hours after treatment (Nardou et al., 2019).

### 4.3.2. Humans

No in-human studies have been published on the post-acute effect of MDMA on functional plasticity.

## 5. Genetic and Molecular markers

The changes in structural and functional plasticity induced by psychedelics require the synthesis of specific molecular building blocks of neural networks. Thus, modification in the expression profile of the genes coding for such proteins is necessary for neuroplastic changes to endure. Neuronal activity- and plasticity-relates genes, coding for related proteins, are modulated by psychedelic compounds in the minutes and days following drug exposure (Figure 2). The most studied class of activity-dependent genes are the immediate-early genes (IEGs), whose activation and transcription begins within minutes after neuronal stimulation. Those include factors such as the activity-regulated cytoskeleton-associated protein (Arc), protein c-Fos, and the Homer protein homolog 1 (Homer1a) (Lanahan and Worley, 1998). Plasticity-related genes encode proteins whose expression was found to be required for the functional and structural remodelling of neurons, including factors such as the brain-derived neurotrophic factor (BDNF), ERKs, and post-synaptic density proteins (PSD) (Ehrlich and Josselyn, 2016). Moreover, epigenetic changes that alter the chromatin status and gene accessibility by transcription factors also play an important role in neuroplastic changes induced by exogenous compounds. Those include histone modifications such as those performed by different histone deacetylases (HDACs) (Figure 2)(Broide et al., 2007; Geng et al., 2021). In animals, a wide variety of biological essays can be performed in vitro and in vivo to measure how neuronal molecular pathways are modulated by psychedelics at various levels, from chromatin status to gene transcription, translation, and protein levels. The most common approach is to measure the gene transcription of specific markers by quantification of the mRNA levels. In humans, only indirect measures are available to quantify molecular profiles of neuroplasticity markers. Circulating levels of specific proteins, such as BDNF in the bloodstream, is the most widely used measure.

### 5.1. Ketamine
#### 5.1.1. Preclinical

In vitro, exposure of primary cortical neurons to low and medium concentrations of ketamine resulted in significant increases in BDNF release acutely and after 1 hour (Lepack et al., 2016, 2014). Ketamine also produces dose-dependent and time-dependent epigenetic changes, such as the stimulation of HDAC5 phosphorylation and its nuclear export in rat-derived hippocampal neurons. Peak phosphorylation was achieved with medium concentrations at 3-6 hours following treatment with a return to baseline after 24 hours. Histone modification was followed by enhanced transcription of the myocyte enhancer factor 2 (MEF2), and subsequent activation of its target genes regulating neuronal structural and functional plasticity (Choi et al., 2015).

In vivo, significant increases in BDNF protein levels have been found in the hippocampus and PFC following administration of medium doses of ketamine in mice immediately and at 1 and 30 days, with no changes following high doses (Kim and Monteggia, 2020; Viana et al., 2020). The antidepressant-like effects produced by a single medium dose of ketamine did not occur in BDNF and eEF2 knock-out mice (Autry et al., 2011; Nosyreva et al., 2013) and synaptogenesis increase was not seen in the mPFC of a mouse models with a BDNF Val66Met low-functioning polymorphism (Liu et al., 2012). In rats, changes in the expression of Arc gene expression and protein levels of Arc, CREB, phospho-CREB, ERK, and phospho-ERK were detected in the hippocampus 10 min and 24 hours after a single medium dose of ketamine (Shi et al., 2021). Another study found upregulation of Arc, Homer1a, and c-Fos transcription in rat's cortical regions 90 minutes after injection (de Bartolomeis et al., 2013). In a mouse model of depression, a medium dose of ketamine reversed the reduction of Arc gene expression in the PFC 72 hours after injection (Bagot et al., 2017). High doses of ketamine were found to only produce an increase in the Arc gene and protein expression levels 10 minutes after injection, but not at later time points. Also, chronic administration of various doses of ketamine produced a decrease of Arc gene and protein expression levels after 24 hours (Shi et al., 2021). In the study by Nardou et al., RNA sequencing of the NAc 48 hours and 2 weeks after a single medium dose of ketamine showed that the IEGs cFos, Junb, Arc, and Dusp and several genes coding for components of the extracellular matrix (ECM) were enriched 48 hours after ketamine, when the critical period of social reword learning is re-opened, as compared to 2 weeks, when the period is closed again (Nardou et al., 2023).

#### 5.1.2. Clinical

In humans, it was observed that MDD patients carrying the Val66Met BDNF allele show reduced response to treatment with medium doses of ketamine (Laje et al., 2012). One study showed an increase in peripheral BDNF at 230 minutes post-administration of a single medium dose of ketamine in MDD patients (Duncan et al., 2013). Another study in healthy individuals found that BDNF levels were higher following ketamine compared to placebo at 2 hours, but this was the function of a reduction of BDNF in the placebo group and BDNF did not significantly increase in the ketamine group (Woelfer et al., 2020). Persisting increases at 2 weeks have also been found following six infusions of medium doses of ketamine (Zheng et al., 2021b). There is some evidence that increased BDNF may be isolated to treatment responders only, with evidence of specificity at 4 hours, 1 week and 1 month post infusion (Allen et al., 2015; Haile et al., 2014; Wang et al., 2021). One study found reductions of BDNF 1 week following a single ketamine infusion in bipolar patients (Rybakowski et al., 2013). Many studies reported no change in serum or plasma BDNF levels following ketamine administration of medium doses of ketamine in the hours or days following administration (Allen et al., 2015; Caliman-Fontes et al., 2023; Glue et al., 2020a, 2020b, 2020c; Grunebaum et al., 2017; Jiang et al., 2021; Machado-Vieira et al., 2009; Medeiros et al., 2021, 2021; Zheng et al., 2022, 2020). Other studies have probed peripheral BDNF levels following the use of ketamine as an anesthetic, either during surgery or electroconvulsive therapy (ECT). Three large studies demonstrated increased BDNF in the days following the administration of a high dose of ketamine compared to placebo (Jiang et al., 2016; Liu et al., 2021; Zheng et al., 2021a). Levels equalized with the placebo at day 5 in one study and persisting for a month in another (Liu et al., 2021; Wang et al., 2020). One study found BDNF to significantly increase following a course of ECT plus ketamine compared to ECT plus placebo, but this was not replicated (Carspecken et al., 2018; Zheng et al., 2021a). Results from studies on frequent ketamine users demonstrate similarly conflicting results (Ke et al., 2014; Ricci et al., 2011).

**5.2. Classic Psychedelics**
**5.2.1. Preclinical**

In vitro, treatment with medium doses of LSD acutely increased gene expression of EGR-1/2 and c-Fos in mice cortical primary neuron cultures (González-Maeso et al., 2007).

In vivo, a study found that a medium dose of DOI differentially regulated BDNF gene expression in hippocampus and neocortex 3 hours after dosing. DOI significantly decreased BDNF expression within the dentate gyrus region of the hippocampus, while induced a dose-dependent increase in the parietal, frontal, and temporal cortices, and in the claustrum. The regulation of BDNF by DOI was found to be dose and time-dependent (Vaidya et al., 1997). Another study found an increase in BDNF gene expression in the medial PFC of rats four weeks after chronic treatment with medium doses of LSD (Martin et al., 2014). Studies examining the effect of a single high dose of LSD on IEGs expression found that c-Fos is increased at approximately 1 and 2 hours after administration in several cortical structures in rodents. Increases in the expression of Arc, Sgk, Nor1, egr-1, and egr-2 were detected at 60-90 minutes, with Nor1 persisting up to 5 hours after LSD administration (Frankel and Cunningham, 2002; González-Maeso et al., 2007; Nichols et al., 2003; Nichols and Sanders-Bush, 2002). A study reported that chronic administration of high doses of LSD leads to increases in transcriptomic complexity and long-term enrichment of genes involved with chromatin organization, covalent chromatin modifications, and histone modifications assessed at 28 days after the last dose (Savino and Nichols, 2022). Medium doses of DOI were found to increase the expression of c-Fos in several cortical and subcortical regions at 30 minutes, 90 minutes, 2 hours, and 3 hours after a single administration in rats (Leslie et al., 1993, p. 2; Moreno et al., 2011; Tilakaratne and Friedman, 1996). Other IEGs which were found to be up-regulated by DOI were EGR-4, and tis1 in cortical regions and ngf1c in the hippocampus and cerebellum at 90 minutes, Arc and EGR-2 in cortical regions at 2 hours (Moreno et al., 2011; Pei et al., 2000; Tilakaratne and Friedman, 1996). The gene expression changes of c-Fos, EGR-4, tis1, and ngf1c at 90 minutes were found to be dependent on 5-HT2AR activation (Tilakaratne and Friedman, 1996), while the effects on ERG-2 and c-Fos at 2 hours were dependent on the activity of the mGluR2 (Moreno et al., 2011). A single medium dose of DOI was found to lead to epigenetic changes in acetylation of histone H3K27 in neuronal nuclei of the mouse frontal cortex, a marker for enhanced transcriptional activity. Specifically, half of the enhancers tested exhibited an increase peaking at 48 hours and declining at 7 days. The rest of the enhancers showed a decrease at 24 hours and a recovery in intensity over 48 hours and 7 days (de la Fuente Revenga et al., 2021). Similar to ketamine, administration of a single medium dose of LSD led to transcriptional upregulation of IEGs and ECM-associated genes in the NAc of mice at 48 hours and 2 weeks, when the critical period of social reward learning is re-opened (Nardou et al., 2023)

**5.2.2. Clinical**

In humans, administration of variable low to medium doses of LSD was found to increase the peripheral level of BDNF at specific dosages and produce no change with others when compared to placebo at 5-6 hours after drug intake in healthy volunteers (Holze et al., 2021; Hutten et al., 2021). Other studies showed no significant effect on plasma BDNF concentrations after the single administration of a medium dose of LSD (Holze et al., 2022, 2020). One study found an increase in plasma BDNF 4-7 hours after a medium dose of psilocybin in healthy individuals, while another study found no effect at 4-12 hours post-administration (Becker et al., 2022; Holze et al., 2022). In health subjects, no changes in whole blood expression levels of EGR1-3 genes were observed at 1.5 and 24 hours following a single medium dose of LSD when compared to the placebo group (Dolder et al., 2017). De Alemida et al. found an increase in serum BDNF in both healthy and depressed patients 48 hours following ayahuasca ingestion compared to placebo. Also, the patients treated with ayahuasca, and not with placebo, presented a significant negative correlation between serum BDNF levels and depressive symptoms (de Almeida et al., 2019).

### 5.3. MDMA
#### 5.3.1. Preclinical

In vivo, a single or repeated medium dose of MDMA in rats increased BDNF mRNA expression in the PFC, NAc, and amygdala, but not in the striatum or hippocampus 2 hours after administration in rats (Mouri et al., 2017). Selective regional effects of MDMA on BDNF expression in adult rats were also observed following repeated administration of medium doses of MDMA with significant increases in BDNF expression detected by *in situ*-hybridization in several cortical regions, including frontal and parietal cortices, at both 1 and 7 hours following treatment, but no changes at 24 hours. Notably, after 1 and 7 hours BDNF expression was decreased in the CA3 and DG regions of the hippocampus but increased in the CA1 region at 24 hours (Hemmerle et al., 2012). A study found that a single medium dose of MDMA increased BDNF expression in rat frontal cortex at 24 and 48 hours and hippocampus at 2 and 7 days following treatment (Martínez-Turrillas et al., 2006), while another found downregulation of the BDNF gene in the hippocampus 24 hours after a single dose (Soleimani Asl et al., 2017). Mice injected with a medium dose of MDMA before fear-extinction training showed a persistent and robust reduction in the conditioned fear, an effect paralleled by an increase in amygdala BDNF expression 1 hour after extinction training. However, this effect was not observed in untrained animals. In the same study, direct MDMA infusion into the basolateral amygdala significantly reduced animals' conditioned fear responses 24 hours following extinction training, which was completely abolished by disrupting BDNF signalling (Young et al., 2015). Pre-treatment with a medium dose of MDMA in a mouse model of minimal traumatic brain injury elevated BDNF levels in the cortex at 1 hour post-administration and protected against injury-induced cognitive deficits through a dopamine-dependent mechanism. However, MDMA administration in control animals led to decreased BDNF levels in the mice striatum, at 24 and 72 hours post-administration (Edut et al., 2014). A single medium dose of MDMA increased c-Fos transcription in rats caudate putamen, NAc, and hippocampus, 1 hour following injection. An increase in expression of EGR-1 was found in the caudate putamen, striatum, hippocampus, and PFC, and an increase in EGR-3 was observed in the caudate putamen 1 hour after administration in rats (Salzmann et al., 2003; Shirayama et al., 2000). Consistent results were obtained in mice 2 hours after a single medium dose of MDMA with an increase in the expression of striatal c-Fos, B-Fos, ERG-1, and EGR-2 (Benturquia et al., 2008) and a brain-wide increase in c-Fos expression in rats (Stephenson et al., 1999). Like ketamine and classic psychedelics, also a single medium dose of MDMA enriched expression of IEGs and genes coding for elements of the ECM in mice NAc at 48 hours and 2 weeks concurrently with the re-opening of social reward learning critical period (Nardou et al., 2023)

#### 5.3.2. Clinical

To date, no direct evidence for the effects of MDMA on gene expression levels has been collected in humans.

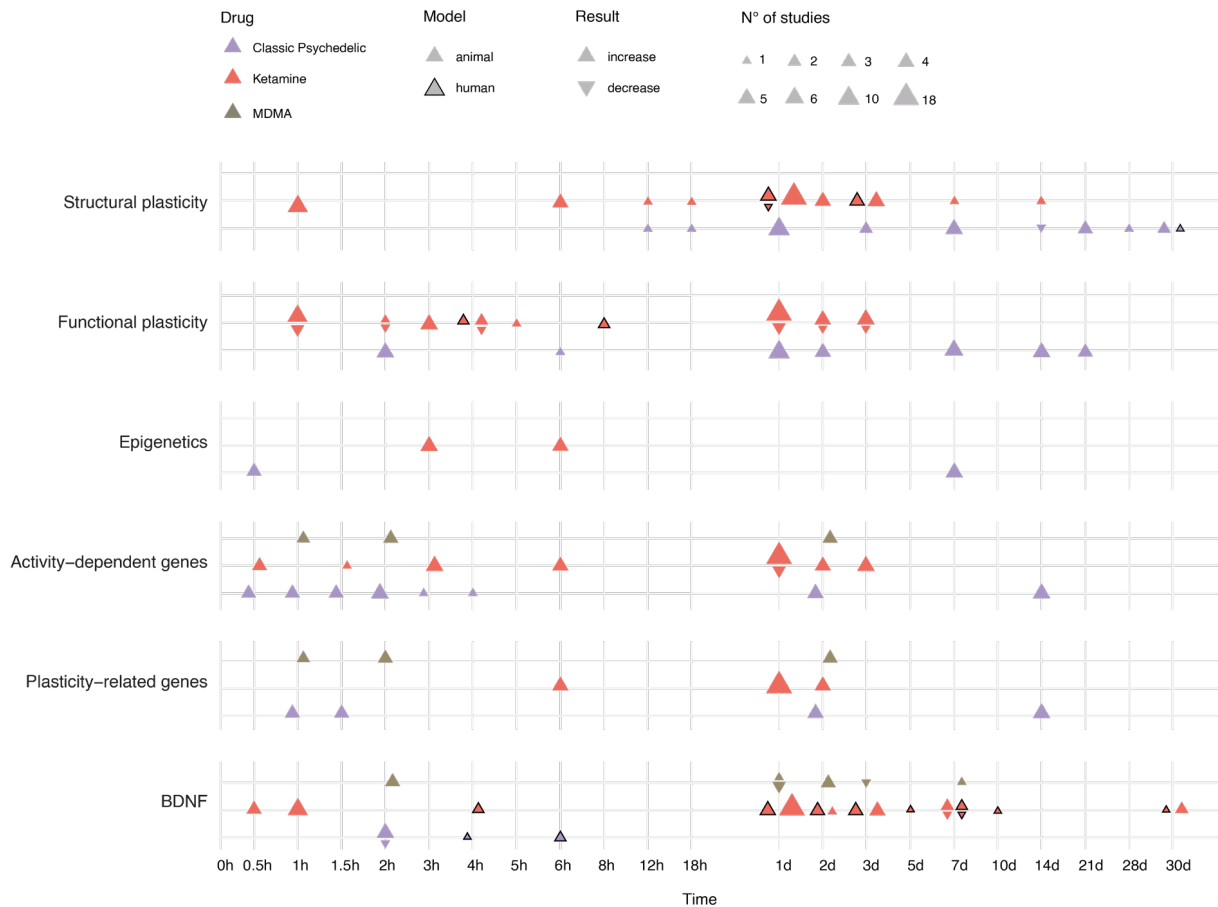

*Figure 3: Post-acute neuroplastic effects of psychedelics. Only the findings relative to the neuroplastic effects induced by a single medium dose of a classic psychedelic, ketamine, or MDMA are shown.*

### 6. Discussion

The results presented on the effects of psychedelics on structural, functional, and molecular markers of neuroplasticity demonstrate that these compounds exert a complex and multifaceted impact on neuroplasticity, producing significant long-lasting biological changes in the organism with behavioral correlates and hence implications for human therapy. In Figure 3, only the results coming from studies investigating the neuroplastic changes induced by a single medium dose of psychedelic compounds are summarised.

On the preclinical side, a framework is emerging to account for the sub-acute effect of a single therapeutically meaningful dose of a psychedelic on neuroplasticity. The acute pharmacological action of psychedelics on neurotransmission would open a state of increased meta-plasticity, for a limited amount of hours, where higher sensitivity to environmental stimuli causes hyper-plastic modifications lasting for weeks or months in a compound-specific way. Such modifications are sustained by up-regulation of molecular and genetic factors and seem to mediate the reduction of behavioral symptoms of mood disorders as reproduced in animal models. In addition, ketamine and classic psychedelics were found to promote adult hippocampal neurogenesis, another mechanism associated with improvements in the symptomatology of mood disorders. The evidence for hyper-plastic modifications (i.e., neuro- and neuritogenesis) following psychedelics exposure is well-characterized both in vivo and in vitro, with increases in dendritic complexity, number of spines, and synapses emerging within a day and lasting for weeks, in the case of ketamine, and months, for the classic psychedelics. In vivo, the structural modifications are mostly defined for neurons within the PFC in the form of increased spinogenesis, where an association has been characterized between neuronal hypertrophy and the reduction of depression-like and anxiety-like animal behaviors. The sustained up-regulation of a molecular pathway involving AMPAR-BDNF-TrkB-mTOR cycling seems to underline the hyper-plastic modifications for ketamine as well as the classic psychedelics. In the case of ketamine, the cycling is initiated acutely via the alteration of glutamatergic neurotransmission, whereas for the classic psychedelics, the activity of 5-HT2ARs seems to be required (Figure 1). Another mechanism common to ketamine and the classic psychedelics to mediate structural plasticity might involve direct allosteric binding to TrkB to induce a meta-plastic state of increased sensitivity to endogenous BDNF. However, this mechanism appears to be independent of 5-HT2AR activity, thus contradicting previous results obtained with the classic psychedelics. Further studies are needed to elucidate the controversy, possibly exploring also the interaction between these receptors (Ilchibaeva et al., 2022). Interestingly, the TrkB pathway appears to be a shared mechanism among classic psychedelics, ketamine, and SSRIs, though with varying degrees of efficacy. Notably, classic psychedelics and ketamine demonstrate greater stability in TrkB binding compared to SSRIs, aligning with their higher potency in inducing neuroplastic modifications.

The evidence for a heightened state of both functional and structural meta-plasticity induced by psychedelics converges from other lines of research. Ketamine exerts a well-replicated effect in hippocampal neurons where the thresholds to undergo stimulation-induced functional modifications, such as short-term and long-term plasticity, and the membrane resting-state potentials are modulated within a few hours and up to a few days after drug administration, depending on dose and stimulation paradigm. Evidence on a sub-acute modulatory effect of functional neural properties exists also for the classic psychedelics, even though with less established paradigms and less characterized behavioral implications than ketamine. Recently, it was observed that both classic and non-classic psychedelics modulate the thresholds to undergo oxytocin-mediated long-term plasticity in rats, an effect associated with the re-opening of a critical developmental period of social reward learning (Nardou et al., 2023, 2019). It was shown that the duration of such effects correlated with the duration of the subjective effects of the different psychedelics, with ketamine showing the shortest durations, followed by psilocybin and MDMA, and then LSD, with the longest duration. This research highlight the importance of taking into account the pharmacodynamic properties of the different psychedelics when considering their neuroplastic and behavioural effects, even though similar data with the short-lasting psychedelics N,N-DMT and 5-Meo-DMT are lacking. Ketamine was also shown to re-open a critical period for functional recovery after monocular deprivation in the visual cortex, which was proposed to be underlined by the meta-plastic increased sensitivity to endogenous BDNF mediated by allosteric binding to TrkB (Cannarozzo et al., 2023). Interestingly, the effects of psychedelics on re-opening critical periods seem to involve extra-neuronal structures, such as relaxation of the perineural nets and the ECM. The convergence of serotonergic psychedelics (the classics and MDMA) with SSRIs in mediating neuroplasticity is particularly interesting in the context of the developmental role of 5-HT in modulating neuroplastic processes. Common pathways of these medications, such as the TrkB and the perineural remodelling, support recent evidence suggesting a critical role of 5-HT in the maturation of neurons (Ogelman et al., 2024), a mechanism proposed to be evolutionary relevant for the expansion of the human cortex (Luppi et al., 2024). In the study by Ogelman et al., the maturation of excitatory synapse within the PFC during development was found to be mediated by 5-HT release on single spines, leading to LTP via 5-HT2A, which in turn increased the long-term survival of newly formed spines, an effect recapitulated by chronic SSRIs treatment (Ogelman et al., 2024). A similar structural meta-plasticity was observed following ketamine as measured via an increase in glutamate-evoked spinogenesis in the PFC for a narrow window of a few hours, providing evidence for the continuity between a temporary state of heightened sensitivity to environmental stimuli and subsequent neural growth (Wu et al., 2021). Altogether, this evidence suggests a general mechanism of a temporal cycle between functional and structural plasticity, mediated by an intermediary process. This cycle begins with a stage of increased meta-plasticity, which promotes hyper-plastic structural modifications. During this phase, the relaxation of extracellular structures, triggered by functional changes in neuronal properties, facilitates neuronal growth and morphological reorganization. This mechanism, active during development to support neuronal maturation, may be harnessed by antidepressants to induce long-term modifications with therapeutic potential.

Of note, the effect of MDMA on neuroplasticity with a single therapeutically meaningful dose is scarce, mostly due to the lack of appropriate design and dosing regimens adopted in the literature. Using high doses, chronic exposure to the drug, or administration during the early stages of development, it was shown that the non-classic psychedelics MDMA and ketamine might have detrimental effects on neuroplasticity. Moving forward, it will be of utmost importance for preclinical research to replicate and compare findings across the different compounds and paradigms.

In humans, translating the evidence on neuroplasticity induced by psychedelics has proven to be challenging. The few available studies attempting to measure structural plasticity following exposure to a psychedelic have failed to produce consistent results. However, some research has successfully probed the heightened meta-plastic state and altered functional neural properties in the hours following exposure to psychedelics with paradigms of non-invasive sensory stimulation. Mixed results were also obtained with regard to genetic and molecular modifications induced by the administration of psychedelic compounds, likely due to technological limitations in the measurements. Therefore, a gap exists in the translation to humans of the working model emerging from the preclinical literature. This discrepancy likely arises from the limited amount of studies and the inherent lack of sensitivity and power of most of the non-invasive imaging methods adopted so far. Emerging methods and techniques offer new opportunities to investigate the sub-acute effects of psychedelics on neuroplasticity in humans. Comparative studies using psychedelics other than ketamine are needed, particularly with novel PET radioligands such as [$^{11}$C]-UCBJ, and the yet-to-be-explored [$^{18}$F]-BCPP-EF and [$^{11}$C]-SA-4503 (Mansur et al., 2019). Additionally, high-resolution structural MRI and DTI for measuring mean diffusivity within grey matter after administering classic psychedelics remain underexplored. Systematic use of sensory stimulation paradigms to examine cortical excitability and meta-plasticity induced by psychedelics would be essential for determining the efficacy and time frame of compound-specific effects (Lyons et al., 2024). Models incorporating sensory deprivation and functional reorganization could also provide valuable insights into psychedelic mechanisms of action. Crucially, the application of non-invasive brain stimulation techniques (NIBS) such as transcranial magnetic, electrical, and ultrasound stimulations in the window of heightened plasticity induced by psychedelics is expected to produce exciting results. This latter approach represents a particularly promising avenue for interventional psychiatry, with the double aim of better characterizing the effects of psychedelics in humans and exploring the synergy between psychedelics and NIBS. Lastly, the combination of the different methods in a multimodal fashion will be critical to capture the complex and dynamic nature of the neuroplasticity process in humans (Lyons et al., 2024). Indeed, while preclinical research provides valuable insights in the mechanism of action of psychedelics, it has limited power in addressing fundamental questions regarding the therapeutic process. Especially in the case of mood disorders, the validity of animal models is confined to the recapitulation of isolated symptoms, which are tested with too unspecific behavioural assays, and cannot investigate the subjective aspect of those illnesses. It is therefore necessary to capture the biological effects of psychedelics in clinical research to appropriately test the hypothesis of neuroplasticity as a potential biomarker.

While a plasticity-promoting effect induced by antidepressants might be necessary to allow for change to take place, the direction of such change would be critically dependent on the state and context of the organism. The concepts of state (i.e., set) and contextual (i.e., setting) dependency, already existing in the tradition of psychedelic research and psychedelic-assisted psychotherapy, now find a neurobiological counterpart in the phenomena of meta-plasticity leading to hyper-plastic modifications. In this way, value is given to extra-pharmacological factors in allowing for therapeutic process to take place, but critical questions remain to be answered. What is the therapeutical value of a potent neuroplastic enhancer without psychological support? What is the role of specific subjective experiences elicited by psychedelics on neuroplasticity and healing? How can we effectively exploit a state of heightened plasticity with targeted non-pharmacological interventions? What is the causal link between acute and sub-acute neuropsychological modifications induced by different psychedelics?

**Supplementary material**

Supplementary table 1 summarises the main results of all studies reviewed in the manuscript.

**Acknowledgements**

The authors wish to acknowledge Leor Roseman, Chris Timmerman, David Reydellet, Brandon Weiss, Mamas Pipis, and Vito Palmisano for their invaluable discussions and scientific contributions to this work

**Funding sources**

The authors declare that no specific funding was received for preparing this manuscript.

**Declaration of generative AI and AI-assisted technologies**

During the preparation of this work the authors used ChatGPT in order to improve language and readability. After using this tool, the authors reviewed and edited the content as needed and take full responsibility for the content of the publication.